\documentclass[pre,showpacs,twocolumn,floats,epsf]{revtex4}
\usepackage{graphicx}
\usepackage{dcolumn}
\usepackage{epsf}
\usepackage{amsmath}
\usepackage{amssymb}

\DeclareSymbolFont{AMSb}{U}{msb}{m}{n}
\DeclareMathSymbol{\N}{\mathbin}{AMSb}{"4E}
\DeclareMathSymbol{\Z}{\mathbin}{AMSb}{"5A}
\DeclareMathSymbol{\R}{\mathbin}{AMSb}{"52}
\DeclareMathSymbol{\Q}{\mathbin}{AMSb}{"51}
\DeclareMathSymbol{\I}{\mathbin}{AMSb}{"49}
\DeclareMathSymbol{\C}{\mathbin}{AMSb}{"43}

\def\openone{\leavevmode\hbox{\small1\kern-3.3pt\normalsize1}}
\newcommand{\la}{\langle}
\newcommand{\ra}{\rangle}

\newcommand{\be}{\begin{equation}}
\newcommand{\ee}{\end{equation}}
\newcommand{\bea}{\begin{eqnarray}}
\newcommand{\eea}{\end{eqnarray}}

\newcommand{\ket}[1]{|#1\rangle}

\newcommand{\unit}[1]{~\text{#1}}
\newcommand{\unitmicro}[1]{~\mu\text{#1}}

\begin{document}

\title{Employing trapped cold ions to verify the quantum Jarzynski equality}
\author{Gerhard Huber and Ferdinand Schmidt-Kaler\\
{\it Institut f\"ur Quanten-Informationsverarbeitung, Universit\"at Ulm,}
 {\it Albert-Einstein-Allee 11, D-89069 Ulm, Germany}\\
Sebastian Deffner and  Eric Lutz\\
{\it Department of Physics, University of Augsburg, D-86135 Augsburg, Germany}}
\date{\today}

\begin{abstract}
We propose a scheme to investigate the nonequilibrium work distribution of a quantum particle under
well controlled transformations of the external potential, exploiting the versatility
of a single ion in a  segmented linear Paul trap. We describe in detail how the
motional quantum state of a single ion can be prepared, manipulated and finally read
out to fully determine the free energy difference in both harmonic and anharmonic potentials.
 Uniquely to our system, we show how an
ion may be immersed in an engineered laser--field reservoir. Trapped ions therefore represent an
ideal tool for investigating the Jarzynski equality in open and closed quantum systems.
\end{abstract}
\pacs{05.30.-d, 05.70.Ln, 37.10.-x}


\maketitle

Nonequilibrium phenomena at the nanoscale are dominated by fluctuations and by
quantum effects. The interplay of nonequilibrium thermodynamics and quantum theory is 
hence of fundamental interest.
 Only a decade ago Jarzynski
published  a major discovery in {\it classical} nonequilibrium physics \cite{jar97},
relating the free energy difference $\Delta F$ after a given transformation to
the  probability distribution of the total work $W$ done on the system:
\begin{equation}
    \label{eq1} \Delta F = - kT \ \ln \la e^{-W/kT}\ra \ ,
\end{equation}
where $\la e^{-W/kT }\ra = \int dW e^{-W/kT} \,P(W)$ is the average exponentiated work
 and $k$ denotes the Boltzmann constant.
This remarkable equality highlights the role of work fluctuations and provides a
generic way of computing the free energy difference for any transformations,
quasistatic or not, once the work distribution $P(W)$ is known. Most importantly, the
Jarzynski relation allows to determine $\Delta F$ even in the case of arbitrarily fast
transformations, when irreversible thermodynamics is not applicable. Prior to the
discovery of Eq.~(\ref{eq1}), the determination of the free energy difference in such
far from equilibrium conditions was not believed to be possible \cite{bus05}. Recently,
the classical Jarzynski equality and its generalization by Crooks \cite{cro99} have
been successfully tested in single--molecule stretching experiments \cite{lip02,col05}.
Later, the work distribution was recovered from repeated measurements of
the mechanical work done on a colloidal particle in liquid solution \cite{blic06}.

The situation is much different at the {\it quantum} level.
So
far only studied theoretically, the Jarzynski equation holds in its classical form for
closed quantum systems \cite{muk03}, while quantum corrections appear in the case of
open systems due to the coupling  to an external reservoir \cite{che04}. Further
difficulties arise when considering the quantum--mechanical nature of work \cite{all05}
and the question of how to measure it experimentally. It is evident that the classical
definition of work as  force times displacement cannot simply be
taken over unmodified. It has recently been established that work is actually not an
observable in the usual sense, as it is not given by an expectation value of some
Hermitian work operator, but rather by a time--ordered correlation function
\cite{tal07a}. On the other hand, the problem of how to determine quantum work still
remains unsolved, explaining the absence of an experimental verification of the
Jarzynski equality in the quantum domain.

We show in this Letter how to experimentally measure nonequilibrium work using a single
ion in a segmented linear Paul trap. A unique property of ion traps is  the possibility
to study the quantum Jarzynski equality, as well as the quantum Crooks relation
\cite{tal07b}, for systems that are either  isolated or coupled to tailored quantum
environments  using  reservoir engineering \cite{poy96,mya00}. Single ions trapped in
radiofrequency Paul traps are quantum nanosystems with remarkable properties. They can
be laser cooled to very low temperatures, reaching to the motional ground state in the
potential. Arbitrary  quantum states can be prepared, manipulated and measured with
high fidelity \cite{haff05,leib05}. Using the so--called electron shelving method, the
quantum state is revealed with close to unity detection efficiency. The use of a
segmented trap further allows for engineering a vast variety of time--dependent
trapping potentials and hence the implementation of different model Hamiltonians.  In
the following, we generalize the detection methods for the motional state
\cite{esc95,app98} in order to realize an efficient filter for vibrational number
states. We show that trapped ions are not only  good candidates for quantum computing,
but may also allow to experimentally approach the emergent field of \mbox{quantum
thermodynamics}.

{\it Quantum Jarzynski equality.} We begin by considering an isolated quantum system
whose time--dependent Hamiltonian is varied  from an initial
value  $H(0)$ to a final value $H(\tau)$. We denote by $\phi_n^t$ and $E_n^t$ the
respective eigenfunctions and  eigenvalues  of the Hamiltonian $H(t)$ at any given time
$t$. We further assume that the system is initially thermalized at temperature $T$. The
free energy difference $\Delta F$ between  final and initial state is then
given by the Jarzynski equality, Eq.~(\ref{eq1}). The probability distribution of the
random work $W$ is given by \cite{tal07a},
 \be \label{eq2}
P(W)=\sum\limits_{m,n}\,\delta[W-(E_{m}^{\tau }-E_{n}^{0})]\,P_{m,n}^{\tau}\, P^0_{n} \
, \ee where $P_n^0=(1/Z_0) \hspace{1.5mm}\exp(-E_n^0/kT)$ is the initial (thermal)
occupation probability and $P_{m,n}^{\tau}$ are the transition probabilities between
initial and final states $n$ and $m$, \be \label{eq3} P_{m,n}^{\tau}=\left|\int dx_0\int
dx \, {\phi_m^\tau}^*(x)U(x,x_0;\tau)\phi^0_n(x_0)\right|^2 \ . \ee Here $ U(x,x_0;\tau)$
is the propagator of the quantum system. The physical meaning of Eq.~(\ref{eq2}) is
clear: the total work done during a given transformation of the Hamiltonian is obtained
from the energy difference between final and initial eigenstates, $E_{m}^{\tau
}-E_{n}^{0}$, averaged over all possible initial and  final states. Equation (\ref{eq2})
shows in addition that the randomness of the work stems from the initial thermal
distribution $P^0_{n}$, and from the quantum nature of transitions between states, as
described by $P_{m,n}^{\tau}$. The origin of work fluctuations is therefore of both
thermodynamical and quantum--mechanical nature. The free energy difference $\Delta F$
can be evaluated for an arbitrary transformation of the Hamiltonian with the help of
the Jarzynski relation, once the work probability density $P(W)$ has been determined. We
next describe a method to realize a quantum nonequilibrium situation for a single trapped ion
in a linear Paul trap and how to measure its corresponding work distribution.

\begin{figure}
\includegraphics[width=6.8cm]{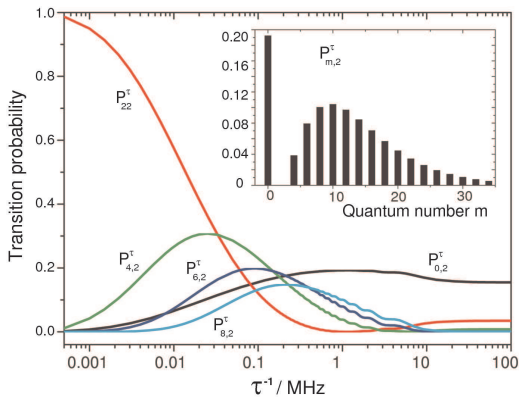}
\caption{Transition probabilities $P^\tau_{m,n}$ for quantum numbers $n=2$, $m=0$ to
$8$ as a function of the inverse switching time $\tau^{-1}$ for a linear transformation of
$\omega^2(t)$ with $\omega(0)=0.1$~MHz and $\omega(\tau)=3.0$~MHz. The inset shows the
transition probabilities $P^\tau_{m,2}$ for a transformation time of $\tau=$1$~\mu$s.
\label{fig:tau_dep_trans_prob_var1}}
\end{figure}

{\it Harmonic ion trap.} Linear Paul traps are characterized by a strong dynamical
confinement in the radial direction ({\it yz} plane) and electrostatically bound in the
axial direction ({\it x} axis). With a radial confinement much stronger than the axial,
we will restrict ourselves to the axial external degree of freedom. Near the center of
the axial potential, the confinement is harmonic and the axial frequency $\omega$ can be varied in time by changing the control voltages \cite{hub08}.
The quantum state of motion along the axial direction can be described by the
Hamiltonian \be \label{eq4} H(t)=\frac{p^2}{2M}+\frac{M}{2}\omega^2(t)\,x^2 \ , \ee
where $M$ is the mass of the ion. For this simple potential, the
nonequilibrium work distribution (\ref{eq2}) can be studied analytically \cite{def08}.
Besides the external, motional degree of freedom, the ion provides  an internal,
electronic level scheme. In our case, we consider a $\Lambda$-system comprising the
ground state S$_{1/2}$ and two excited states P$_{1/2}$ and D$_{5/2}$. The P$_{1/2}$
state rapidly decays into the S$_{1/2}$, thus providing a high spontaneous photon
scatter rate used for fluorescence detection. Laser induced transitions from the ground
to the metastable D$_{5/2}$ state are induced on the narrow quadrupole transition
(linewidth $\Gamma_D\ll \omega(t)$), if the spectral bandwidth of the
S$_{1/2}$-D$_{5/2}$ exciting light field is small compared to the sideband structure.
Coherent laser pulses on this narrow-band optical transition allow to exploit and to
store the motional quantum state information in the internal quantum states.

The experimental {\it measurement protocol} of the  work distribution consists of four consecutive steps:

(I) The ion is first prepared in a thermal state with mean phonon number, $\bar{n}=(\exp(\hbar \omega_0/kT)-1)^{-1}$,  in the
electronic ground state S-level by laser cooling and optical pumping. We
prepare this state deterministically by resolved--sideband laser cooling \cite{schulz08}
into the vibrational ground state $\ket{n=0}$ and subsequently allowing the ion to heat up for a
certain time without laser cooling. As the heating rate of the ion within the trap can
be precisely measured, this procedure is favorable for very low values of $\bar{n}$. An
alternative method, suited for higher values of $\bar{n}$, is Doppler cooling on the
S$_{1/2}$ to P$_{1/2}$ transition. Varying the detuning of the cooling laser from the
atomic resonance results in different thermal states with mean phonon numbers down to
the Doppler limit.

\begin{figure}
\includegraphics[width=6.5cm]{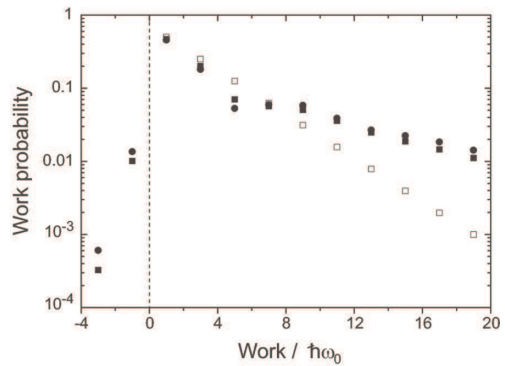}
\caption{Work probability distribution,  Eq.~(\ref{eq2}), for a linear change of the harmonic potential
$\omega^2(t)$ with $\omega(0)=1\unit{MHz}$ and $\omega(\tau)=3\unit{MHz}$, for  $\overline{n}=1$.
Shown are the adiabatic case $\tau\rightarrow\infty$ ($\square$) and
two fast transformations with $\tau=0.1\unitmicro{s}$ ($\blacksquare$) and
$\tau=0.05\unitmicro{s}$ ($\bullet$).
For the latter, deviations from the adiabatic case are clearly visible. Even negative work processes arise: the probability contribution at
$W=-\hbar\omega_0$ mainly stems from the transition $n=2\rightarrow m=0$,
which occurs with probability $P^\tau_{2,0}=10\%$ (for $\tau=0.05\unitmicro{s}$). This
contribution can be readily tested by the number state filter.}
\label{fig:work_distribution}
\end{figure}

(II) In the second step, we measure the initial phonon number $n$ using the filtering
scheme described in detail below. In this way, we determine the initial energy
eigenstate $E_n^0$  (from spectroscopy measurements).

(III) In the third step, we transform the trap potential from an initial value
$\omega(0)$ to a final value $\omega(\tau)$ This changing potential will in general
modify the ion's motional state into a nonequilibrium  state, while its internal,
electronic state remains unaffected. For simplicity, we consider a linear variation of
the confining axial potential $\omega^2(t)$  from $\omega^2(0)$ to $\omega^2(\tau)$.
Figures \ref{fig:tau_dep_trans_prob_var1}  and \ref{fig:work_distribution} show a
numerical evaluation,  based on the results of Ref.~\cite{def08},  of the transition
probabilities (\ref{eq3}) and the  work distribution (\ref{eq2}) for realistic
experimental parameters and different transformation times $\tau$.

(IV) In the last step, we measure the new phonon number $m$ using the filtering scheme
and determine the final energy eigenstate $E_m^\tau$. The distribution of the
nonequilibrium work, $W= E_m^\tau-E_n^0$, Eq.~(\ref{eq2}), is then reconstructed by
repeating the measurement sequence. By evaluating Eq.~(\ref{eq1}) for
adiabatic and  nonadiabatic processes, we can verify the Jarzynski equality.

\begin{figure}
\includegraphics[width=6.2cm]{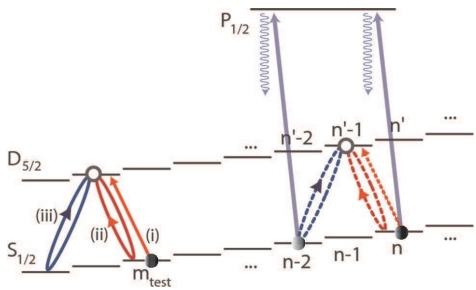}
\caption{Levels and transitions involved in the filtering scheme. Shown are transitions
driven by the initial $\pi$-pulse (i) and the successive $2\pi$-red sideband (ii) and
$2\pi$-blue sideband pulses (iii), respectively. For $|m_\textrm{test}\rangle$, these pulses
induce perfect $\pi$ and $2\pi$-transitions between the metastable $\textrm D_{5/2}$
and the ground state level $\textrm S_{1/2}$ (left side). No fluorescence is observed
when the ion is exposed to resonant light on the S to P transition. For
any other $|n\rangle$, the transfer will be imperfect (dashed, right side) and there
will be  population in the S. Thus, the excitation to the P
level is successful and we observe the emission of fluorescence photons.}
\label{fig:filter_schema}
\end{figure}

\begin{figure}
\includegraphics[width=7.5cm]{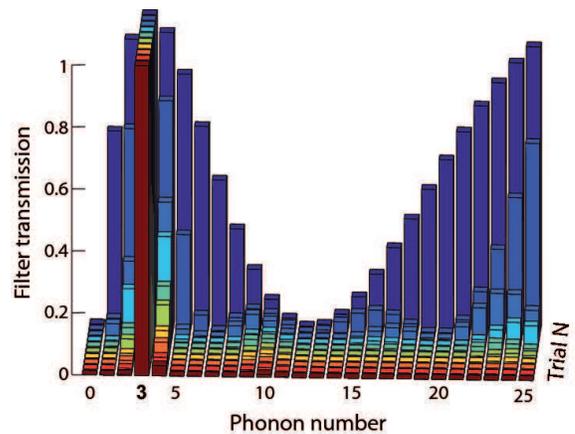}
\caption{Probability for detection of zero--fluorescence, called filter transmission,
after $N=1,2,\dots 10$ pulse/detection trial cycles. With increasing number of trials
(from back to front) the Fock state to be tested for, here $m_\textrm{test}$ = 3, always
remains dark under excitation with resonant light on the S$_{1/2}$ to P$_{1/2}$
transition. All other states show an exponential decrease of zero--fluorescence
detection probability. } \label{fig:filter}
\end{figure}

{\it Filtering scheme.} A sequence of laser pulses on the narrow S to D transition is
applied to the ion, coherently processing its internal and external degrees of freedom
\cite{ben03}. We tailor this pulse sequence such that the ion will  end in the
metastable D$_{5/2}$ state with certainty if the vibrational quantum state was
$\ket{m_\textrm{test}}$. Subsequently, the ion is illuminated with light resonant to the
S$_{1/2}$ to P$_{1/2}$ transition. If we observe no fluorescence, the ion is measured
in the D state. However, for vibrational states different from $|m_\textrm{test}\ra$, the
laser pulse sequence leads to a superposition state, $\alpha \ket{S} + \beta \ket{D}$,
such that there remains a non--vanishing probability $|\beta|^2$ of projecting the
superposition into $\ket{D}$, and thus observing no fluorescence. We therefore
repeat the procedure a few times such that a high quality of the filtering procedure is
ensured. Considering the evolution of the quantum state itself, the influence of the
above sequence reminds of the operating principle of a filter: its projective
'transmission' is unity for a certain input state $|m_\textrm{test}\ra$ and zero otherwise. We
adapt the laser pulse sequence timing to reach all relevant eigenstates
$\ket{n}$ and $\ket{m}$ with the filter.

The crucial requirement for a well--suited filter procedure is to ensure the non--zero
fluorescence outcome for all states but $|m_\textrm{test}\ra$. It is sufficient to design the
number state filter to have high suppression factors in a vicinity of $|m_\textrm{test}\ra$,
since $P^\tau_{m,n}$ is rapidly vanishing for high values of $|m-n|$. Varying the
duration of the pulses, we use the following procedure, see
Fig.~\ref{fig:filter_schema}. Starting from state $|S,m_\textrm{test}\rangle$, we apply
a $\pi$-pulse on the first red sideband leaving the ion in
$|D,m_\textrm{test}-1\rangle$. As the Rabi frequency $\Omega_{n,n'}$ between vibrational
states $n$ and $n'$ depends on both initial $n$ and final $n'$, the laser pulse does
not completely transfer other vibrational states to the D state. When we expose
the ion to resonant light on the S to P transition, zero--fluorescence
is observed if the ion was in $|m_\textrm{test}\ra$, but for other vibrational states there is
a certain probability to observe fluorescence. To sharpen the discrimination, we apply
$2\pi$-pulses on the red and blue sideband, respectively, interleaved by a fluorescence
detection trial after each pulse. Again, the $2\pi$-pulses and detections leave the ion
in the dark state $|D,m_\textrm{test}-1\rangle$, but yield a non-zero fluorescence signal for
all other states. This resembles so--called trapping states which have been
investigated in cavity QED experiments 
\cite{weid99}. This probability for zero--fluorescence detection decreases
exponentially with the number $N$ of pulse/detection runs, as shown in
Fig.~\ref{fig:filter}, for a wide range of states $m_\textrm{test}\geq 3$. After $N=10$ runs,
all probabilities but for $|m_\textrm{test}\ra$ drop below $5~\%$. As no coherence is
remaining after each detection interval, the scheme has modest requirements on the
phase stability. The driving laser needs preserving phase only for one single
$2\pi$-pulse, but not during the entire filter sequence. The two lowest vibrational
states are treated even simpler: For $m_\textrm{test}$=0, a $\pi$-pulse on the carrier
transition brings the ion into $|D,0\rangle$. Successive red sideband $\pi$-pulses do
not affect this state, but fluorescence is observed with non--zero probability for all
other states. This scheme has been proposed for stochastic cooling \cite{esc95,app98}.
For $m_\textrm{test}$=1, the carrier pulse is simply replaced by a red sideband $\pi$-pulse
$\ket{\textrm S,1}\rightarrow\ket{\textrm D,0}$. Then the procedure continues as for
$m_\textrm{test}$=0. The length of the pulses is specific for each choice of $m_\textrm{test}$,
varying the pulse allows to access measurements over a wide range of vibrational
levels.

To estimate the time taken by one experimental  cycle from preparation to detection,
we assume a few $10\unitmicro{s}$ for sideband pulses and a few hundred $\mu$s
fluorescence detection time; one cycle with multiple filtering iterations will then take
less than $ 10\unit{ms}$, short compared to the lifetime of the D-state ($1.2\unit{s}$
for $^{40}\textrm{Ca}^+)$. To also assure that unwanted dissipative effects do not
introduce errors, the trap's heating rate needs to be smaller than 1 phonon within the cycle time. Traps with much lower  rates have been reported. The
statistical error of the values $P^\tau_{m,n}$ is further reduced by repeating the measurement
sequence.

{\it Designing the bath properties and the potential shape}. As discussed before,
single trapped ions are highly isolated from external reservoirs. However, it has been
shown theoretically \cite{poy96}, and also in first experiments \cite{mya00}, that it
is possible to introduce a coupling between the ion and an artificially laser-induced
bath. The variation of laser frequencies and intensities allows one to engineer the
coupling and select the master equation describing the motion of the ion. Here, the
interaction is mediated by sideband transitions between the S$_{1/2}$ and D$_{5/2}$
level, see Fig.~\ref{fig:filter_schema}. For example, a zero--temperature reservoir can
be implemented by a light field tuned to the (cooling) red sideband transition. A large
variety of other tailored reservoirs, such as squeezed baths, can be generated as well.
Within the framework of this proposal, it is therefore possible to investigate
nonequilibrium transformations of open systems with tailored couplings between system
and reservoir. In particular, the distribution of the heat exchanged with a reservoir
can be determined using  the same measurement protocol by keeping the trap frequency
constant, that is, performing no work.

Exploiting the flexibility provided by a segmented trap design, it is moreover feasible
to investigate anharmonic trapping potentials. Especially in situation of a
nonadiabatic transport along the segments of the trap \cite{hub08}, the ion is shifted
out of the harmonic center of the electric potential and experiences nonharmonic
potential contributions \cite{schulz08}. For future work, one might include forces by
laser light on the ion, which depend on its internal electronic state, investigating
the influence of quantum thermodynamics on qubit gate operations \cite{hen07}.

In conclusion, we have shown how the quantum Jarzynski equality can be experimentally
investigated using a single ion in the time--varying electrical potential of a Paul
trap, for both open and closed quantum systems. Our proposal is based on the
state--of--the--art in many laboratories working with single trapped ions and uses
realistic parameters. Experiments with such a device would allow to shine more light on
the amazing interplay of quantum mechanics and thermodynamics.

 This work was supported by the German science foundation within the
SFB/TRR-21 and the Emmy Noether Program (Contract LU1382/1-1), by the European
commission within MICROTRAP (Contract 517675) and EMALI (Contract MRTN-CT-2006-035369),
and the cluster of excellence Nanosystems Initiative Munich (NIM). We thank W. Schleich
for discussion.

%
%

%
%

\end{document}